\documentclass[a4paper]{article}
\usepackage[utf8]{inputenc}
\usepackage{biblatex}
\usepackage{graphicx}
\usepackage{booktabs}
\usepackage{amsmath}
\title{Autoencoders for Multi-Label Prostate MR Segmentation}
\author{Ard de Gelder\footnote{Ard de Gelder is with the Diagnostic Image Analysis Group, Radboud University Nijmegen Medical Centre (e-mail: arddegelder@solcon.nl)}, Henkjan Huisman}
\date{\today}

\begin{document}
\maketitle

\begin{abstract}
 Organ image segmentation can be improved by implementing prior knowledge about the anatomy. One way of doing this is by training an autoencoder to learn a lowdimensional representation of the segmentation. In this paper, this is applied in multi-label prostate MR segmentation, with some positive results.
\end{abstract}

\section{Introduction}
Prostate cancer is a major cause of cancer mortality among men. Multi-parameter MRI is being used, both in diagnosis and in treatment of prostate cancer. For these purposes, however, segmentation is necessary, which requires a lot of expertise. So automation of this task could greatly benefit clinical practice, possibly even enabling population-wide pre-emptive screening. 
In particular, it would be useful to segment two different zones within the prostate: the transition zone (TZ) and peripheral zone (PZ), since these have differing guidelines for mpMRI diagnosis of cancer. This is quite challenging, especially the border between the two zones is subtle and hard to segment.

Automatic segmentation has greatly improved recently, most notably due to the use of convolutional neural networks like UNet. \cite{cicek} A variant, VNet, has been applied to full prostate segmentation \cite{milletari} and recently a 3D-version of UNet has been used to do multi-zonal prostate segmentation. \cite{germonda}

In cardiac imaging, autoencoders have been used as a way of implementing prior knowledge into neural networks, with some positive results. \cite{oktay} We hypothesize that the same techniques can be used to increase the accuracy of automatic multi-label prostate segmentation.

\begin{figure}
    \centering
    \includegraphics[width=0.45\textwidth]{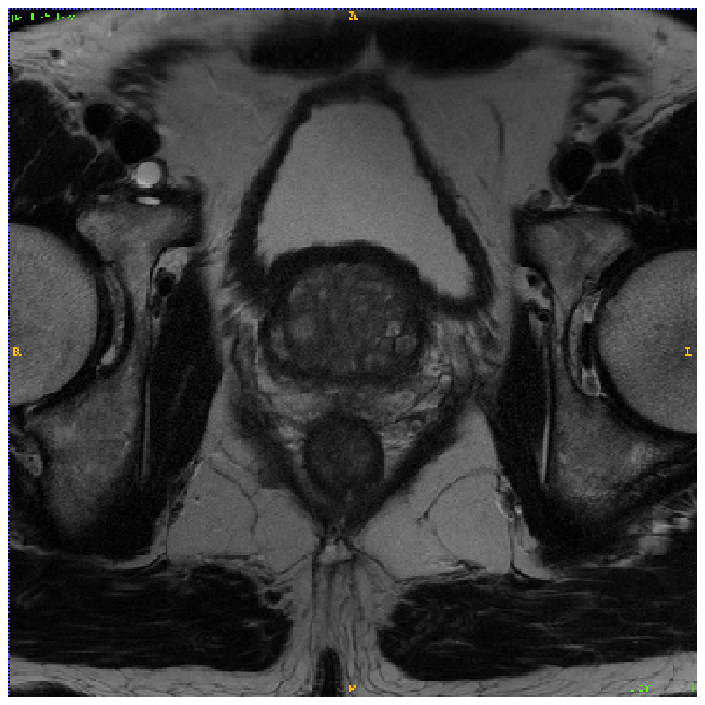}
    \includegraphics[width=0.45\textwidth]{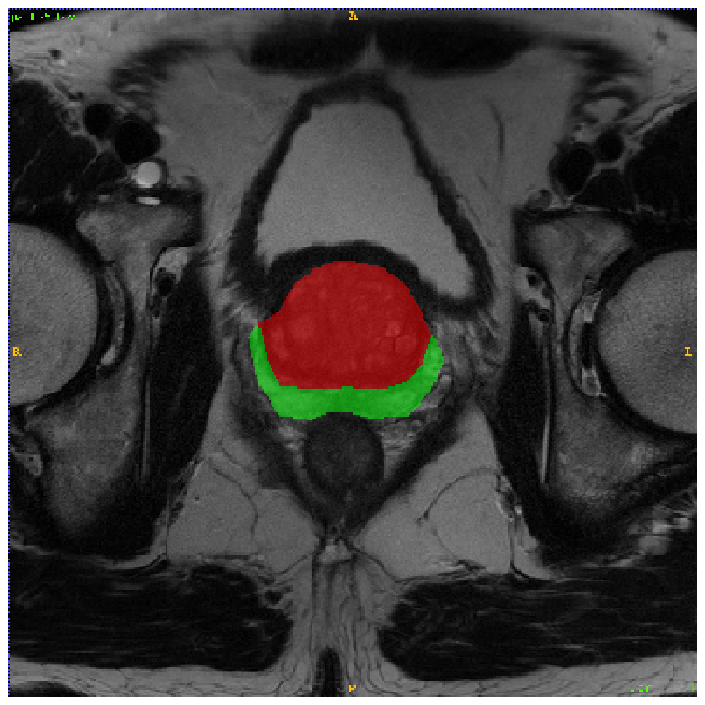}
    \caption{MRI-slice (left) with manual segmentation (right). Red = TZ, Green = PZ}
    \label{manual_seg}
\end{figure}

\section{Dataset}
The used dataset constists of 64 3D T2-weighted MRI volumes of the prostate and surrouding region from the 2016 Detection Archive \cite{dataset}. In these volumes both TZ and PZ are annotated by hand. See for example Figure \ref{manual_seg}. 

The original images are too large to fit in memory, so they are cropped and rescaled, see Table \ref{rescale}.

During training, the data is augmented by small transitions, left-right flips, isotropic expansions, elastic deformations, and rotations.

\begin{table}[h]
    \centering
    \caption{Image rescaling}
    \begin{tabular}{c|c c |c c}
    \toprule
        dimension & original voxels & voxel size & rescaled voxels & voxel size\\
        \midrule
        x & 384 or 640 & 0.5mm & 36 & 3mm\\
        y & 384 or 640 & 0.5mm & 36 & 3mm\\
        z & 19-27 & 3.6mm & 18 & 3.6mm \\
    \bottomrule
    \end{tabular}
    
    \label{rescale}
\end{table}

\section{Methods}
The main network used for segmentation is based on the 3D-UNet architecture, with one modification: to reflect the anisotropicity of the MRIs, some 3D-convolutions were replaced by 2D-convolutions. \cite{cicek, germonda}

In an attempt to improve this, an autoencoder is added.

\subsection{Autoencoder}
An autoencoder is a neural network that consists of two parts: an encoder, that reduces a given segmentation to a lower-dimensional encoding and a decoder, that aims to reconstruct the original segmentation from the encoding as accurately as possible.
\begin{figure}[h]
    \centering
    \includegraphics[width=0.6\textwidth]{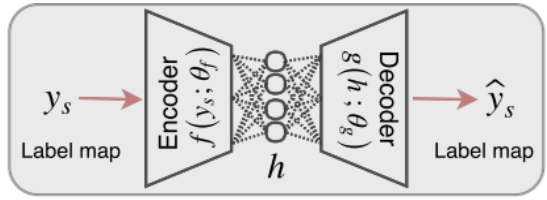}
    \caption{Encoder. Figure from \cite{oktay}}
    \label{encoder_idea}
\end{figure}

Since the size of the encoding is lower than the input, the encoder has to capture the most important features of the data. So this lower-dimensional encoding can be used as a summary of the global properties of a segmentation.

\begin{figure}
    \centering
    \includegraphics[width=0.3\textwidth]{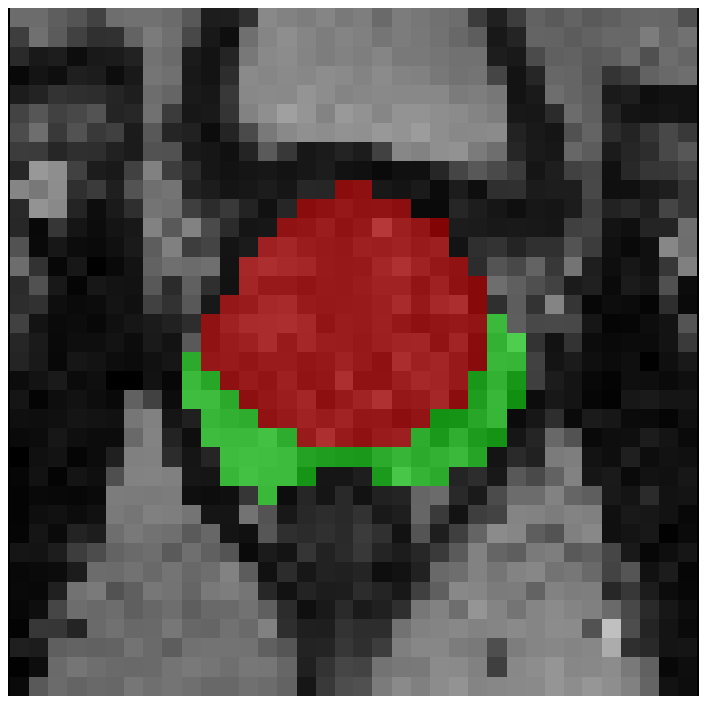}
    \includegraphics[width=0.3\textwidth]{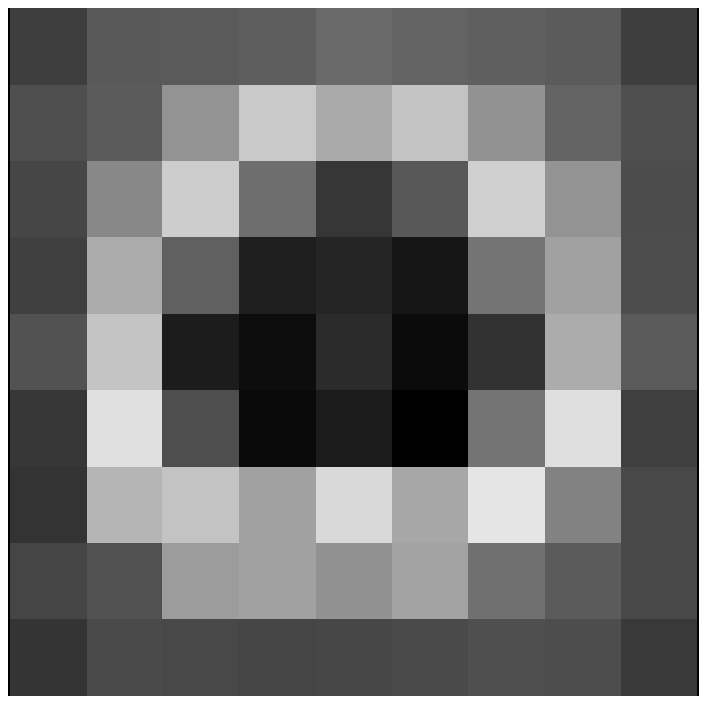}
    \includegraphics[width=0.3\textwidth]{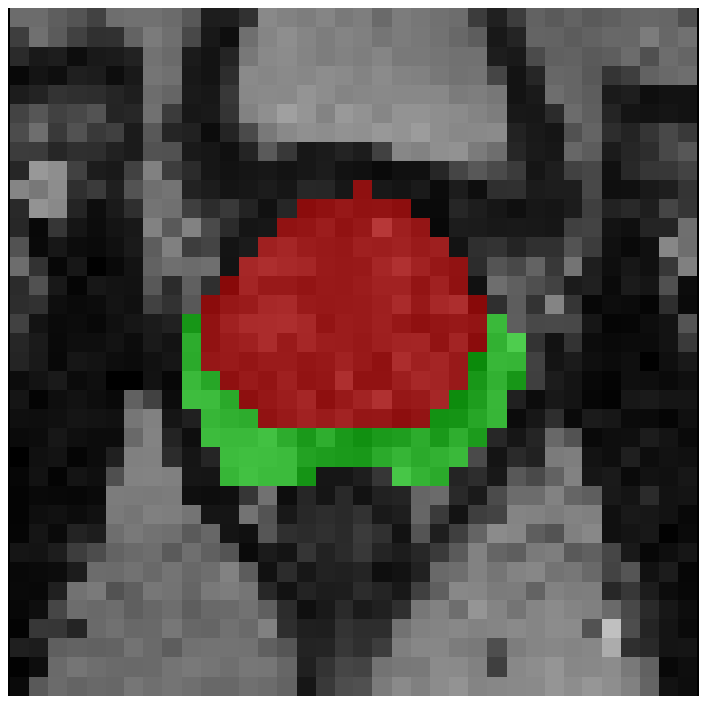}
    \caption{Left: ground truth, center: one encoding slice, right: reconstruction}
    \label{encoder_reconstruction}
\end{figure}

The used autotoencoder is a fully convolutional one that reduces the 36x36x18  segmentation to a 9x9x5 encoding. See for an example Figure \ref{encoder_reconstruction}.

This autoencoder is trained on the manual annotations in the dataset for 100 epochs, using binary crossentropy loss and an Adam optimizer. \\

The main metric used to evaluate performance is the DICE score, given by
$$DICE(P, G) = \frac{2|P \cap G|}{|P| + |G|} $$
where $P$ is the prediction and $G$ the ground truth. \\

The autoencoder could reconstruct the TZ with an average DICE of 0.95 and the PZ with a DICE of 0.85.\\

\begin{figure}
    \centering
    \includegraphics[width=0.9\textwidth]{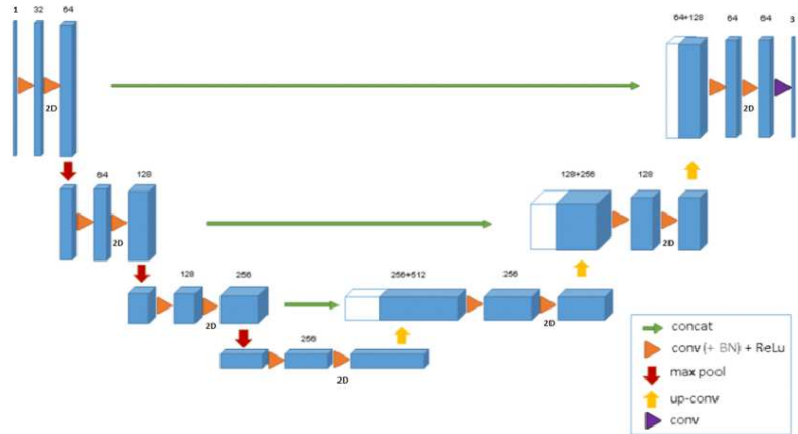}
    \caption{3D-UNet based artitecture. Figure from \cite{germonda}}
    \label{unet_architecture}
\end{figure}

\begin{figure}
    \centering
    \includegraphics[width=0.9\textwidth]{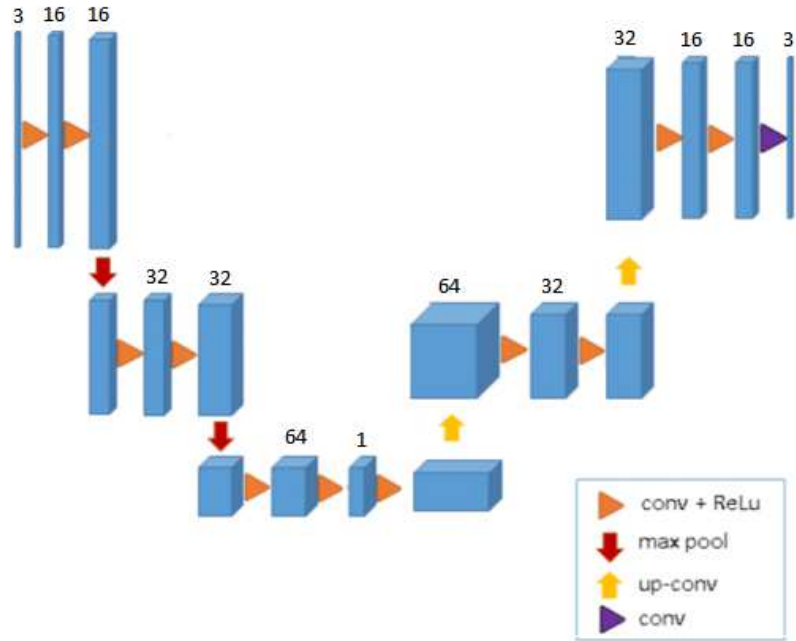}
    \caption{Autoencoder architecture}
    \label{encoder_architecture}
\end{figure}

\subsection{Implementation}
\begin{figure}
    \centering
    \includegraphics[width=0.9\textwidth]{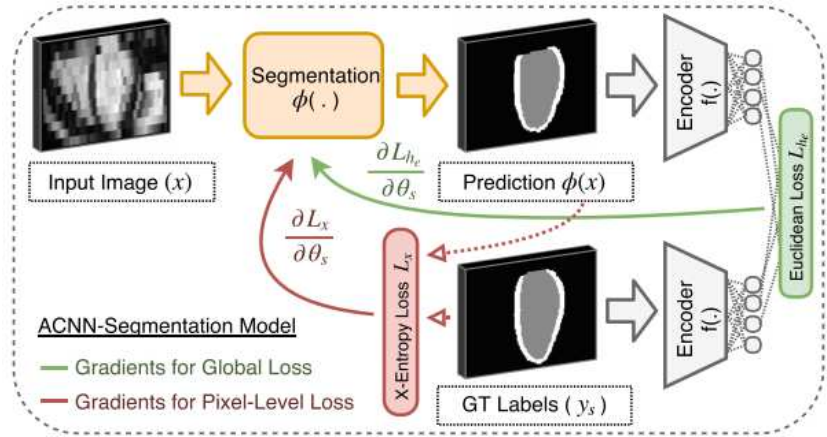}
    \caption{Implementation of the encoder. Figure from \cite{oktay}}
    \label{encoder_implementation}
\end{figure}

During training of the 3D-UNet, the pre-trained encoder is used to add an extra global loss, as seen in Figure \ref{encoder_implementation}. This global loss is added to the pixel-wise loss, where the pixel-wise loss has a weight factor of 1 and the encoder-generated global loss a weight factor of 0.2. \\
The pixel-wise loss is calculated by weighted categorical crossentropy, where the background has weight 1, the TZ weight 2, and the PZ weight 6, in order to compensate for label inbalances.

The 3D-UNet was trained for 300 epochs, using an Adam optimizer, a learning rate of 0.0001 and L2 kernel regulazation.

\section{Results}
First the 3D-UNet was trained without the extra loss provided by the encoder, and the results were compared to a 3D-UNet that was trained with the extra encoder-based loss. 

The 3D-UNet that was trained with the encoder obtained slightly better results.
\begin{table}[h]
\centering
\caption{Segmentation DICE scores}
\label{dice_with_without}
\begin{tabular}{l|cc}
\toprule
   & TZ & PZ \\
\midrule
3D-UNet & 0.85            & 0.60         \\
3D-UNet trained with encoder & 0.85            & 0.67        \\
\bottomrule
\end{tabular}
\end{table}

\begin{figure}
    \centering
    \includegraphics[width=0.3\textwidth]{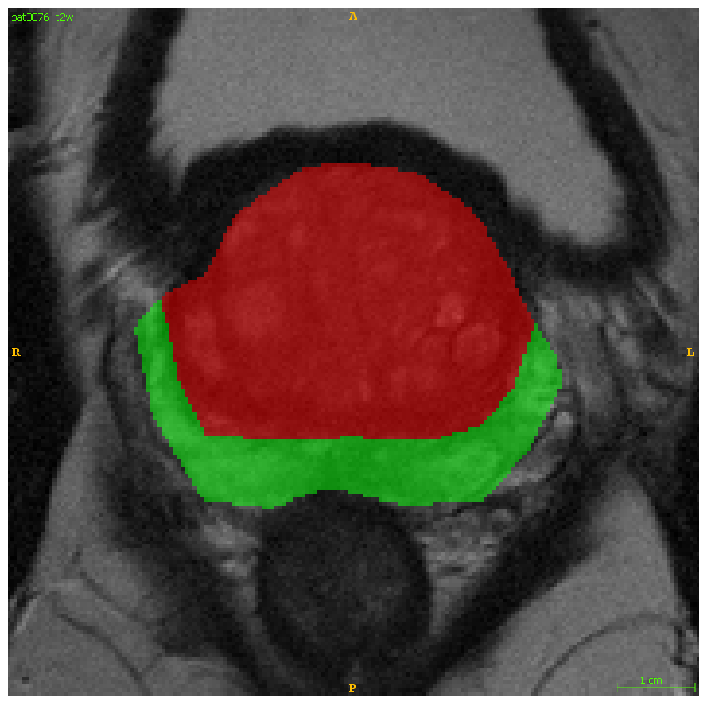}
    \includegraphics[width=0.3\textwidth]{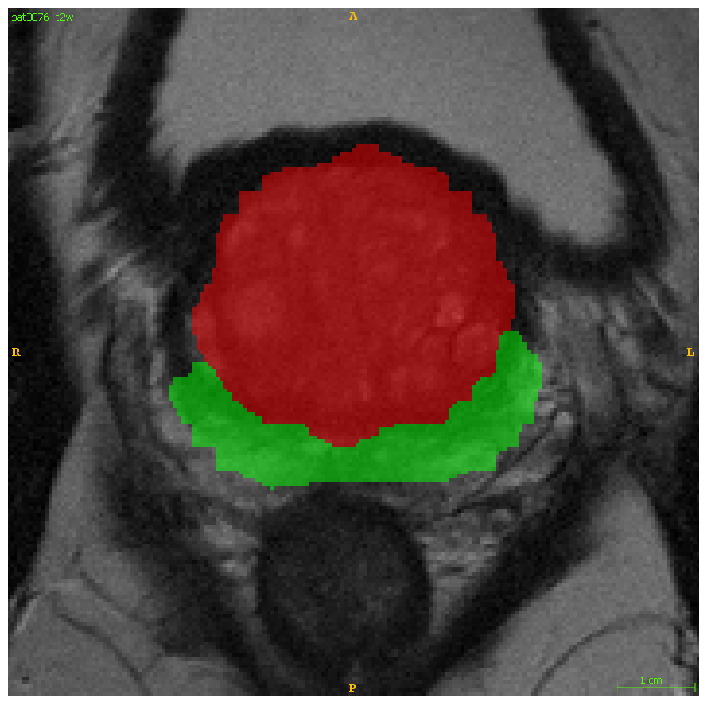}
    \includegraphics[width=0.3\textwidth]{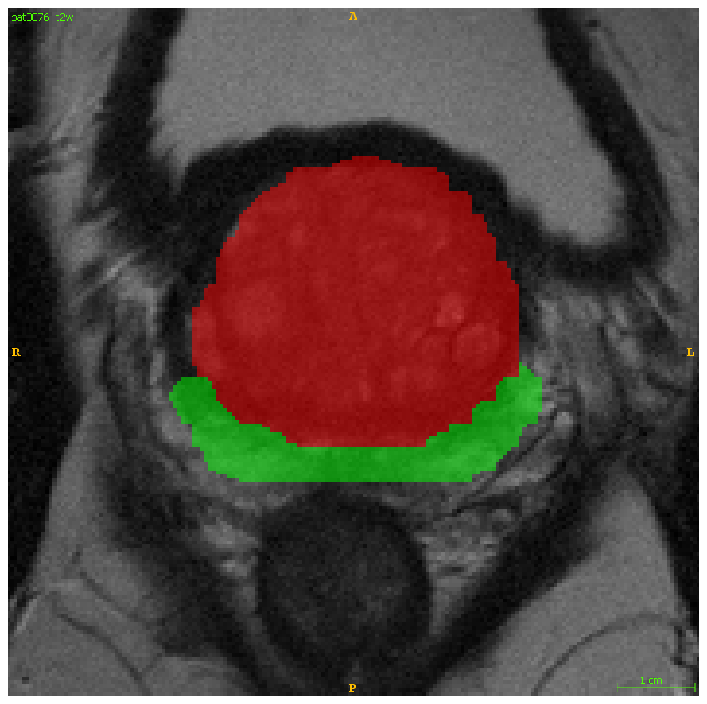}
    \caption{Left: manual segmentation (ground truth), center: segmentation by 3D-UNet trained without encoder, right: segmentation by 3D-UNet trained with encoder}
    \label{gt_with_without_comparison_78}
\end{figure}

\section{Conclusions}
In this work we applied convolutional autoencoders to aid the training of a 3D-UNet in multi-label prostate segmentation. This did increase the segmentation accuracy, but only slightly. \\
One of the reasons the improvement is quite small could be that the image size is already reduced quite significantly before using it in 3D-UNet. It could be studier further whether for larger images, the autoencoder has more impact on performace. \\
Another possible improvement would be gradually decreasing the weight of the encoder-based loss while training the 3D-UNet, since the contribution the encoder-based global loss is largest during the beginning of the training. 

\printbibliography

\end{document}